\pdfoutput=1

\documentclass[11pt]{article}

\usepackage{ACL2023}
\usepackage{amsmath}
\usepackage{newtxmath}

\usepackage{times}
\usepackage{latexsym}
\usepackage{multirow}
\usepackage{booktabs}
\usepackage{graphicx}
\usepackage{makecell}
\usepackage{subfig}  
\usepackage{url}
\usepackage[normalem]{ulem}
\useunder{\uline}{\ul}{}
\usepackage[T1]{fontenc}

\usepackage[utf8]{inputenc}

\usepackage{microtype}

\usepackage{inconsolata}
\usepackage{color}
\newcommand{\tcolor}{\textcolor{black}}
\title{
Understanding Differential Search Index for Text Retrieval}

\author{$\text{Xiaoyang Chen}^{1,2} \text{, Yanjiang Liu}^{1,2} \text{, Ben He}^{1,2*} \text{, Le Sun}^{2*} \text{, Yingfei Sun}^{1*}$\\
  $^{1}\text{University of Chinese Academy of Sciences, Beijing, China}$ \\
  $^{2}\text{Institute of Software, Chinese Academy of Sciences, Beijing, China}$ \\
  \texttt{\{chenxiaoyang19, liuyanjiang22\}@mails.ucas.ac.cn}, \\ 
  \texttt{\{benhe, yfsun\}@ucas.ac.cn} \\
  \texttt{sunle@iscas.ac.cn} \\
  }

\begin{document}
\maketitle
\begin{abstract}
The Differentiable Search Index (DSI) is a novel information retrieval (IR) framework that utilizes a differentiable function to generate a sorted list of document identifiers in response to a given query. 
However, due to the black-box nature of the end-to-end neural architecture, it remains to be understood to what extent DSI possesses the basic indexing and retrieval abilities. To mitigate this gap, in this study, we define and examine three important abilities that a functioning IR framework should possess, namely, exclusivity, completeness, and relevance ordering. Our analytical experimentation shows that while DSI demonstrates proficiency in memorizing the unidirectional mapping from pseudo queries to document identifiers, it falls short in distinguishing relevant documents from random ones, thereby negatively impacting its retrieval effectiveness. To address this issue, we propose a multi-task distillation approach to enhance the retrieval quality without altering the structure of the model and successfully endow it with improved indexing abilities. Through experiments conducted on various datasets, we demonstrate that our proposed method outperforms previous DSI baselines\footnote{The code and data for this work can be found at \url{https://github.com/VerdureChen/Understang_DSI}}.
 
\end{abstract}

\section{Introduction}

Recent advancements in the field of information retrieval (IR) have sparked a growing interest in Differentiable Search Index (DSI)~\citep{DBLP:journals/corr/abs-2202-06991}. Unlike traditional methods, which involve building an index before retrieval~\citep{DBLP:journals/corr/abs-1910-10687, nogueira2019doc2query, DBLP:journals/corr/abs-2010-11386, DBLP:conf/iclr/XiongXLTLBAO21}, DSI and related techniques such as DSI-QG~\citep{DBLP:journals/corr/abs-2206-10128} and NCI~\citep{DBLP:journals/corr/abs-2206-02743} do not rely on external indexes to store data. Instead, these methods map user queries directly to the identifiers (IDs) of the relevant documents, providing a simpler and more efficient retrieval process. This novel autoregressive approach, represented by DSI, has expanded the potential IR applications due to its ease of use, minimal index storage requirements, and end-to-end retrievability.



However, despite the novel retrieval mechanism of DSI, current DSI models still rely on relevance signals of query-passage pairs for training. These models, which map short texts to specific document IDs, do not have an explicit interaction between the query and the document during retrieval, unlike dense retrieval models~\citep{DBLP:journals/corr/abs-2004-12832,DBLP:conf/sigir/HofstatterLYLH21, DBLP:conf/naacl/QuDLLRZDWW21,lin-etal-2021-batch,karpukhin-etal-2020-dense, DBLP:conf/emnlp/GaoC21} and cross-attention rerankers~\citep{DBLP:journals/corr/abs-1910-14424,DBLP:journals/corr/abs-1901-04085,DBLP:conf/emnlp/ZhengHHH0Y20, DBLP:journals/corr/abs-2008-09093, DBLP:journals/corr/abs-2002-10957,DBLP:conf/ecir/ChenHHHSY22}. This training approach and the inherent properties of the model may lead to two problems. First, due to the lack of explicit modeling of inter-document associations and an explicit query-document relevance measurement, the model may only learn a unidirectional mapping from short texts to specific IDs, without understanding how the document is relevant to the query, leading to somewhat random output in the ranking list. Second, to reduce computational complexity, DSI models often simply represent a document by a small number of tokens or pseudo-queries. However, this approach may result in a reduced capacity to differentiate between documents and capture crucial relevant information.

This study aims to deepen the understanding of DSI by evaluating its suitability as an end-to-end model for indexing and retrieval. To achieve this, DSI-QG~\citep{DBLP:journals/corr/abs-2206-10128}, a recent enhancement of DSI using pseudo queries for the model training, is used as a representative model for analysis. Our argument is that a usable index of a non-boolean retrieval model should meet the following three conditions: 1) The document content in the index should have a one-to-one correspondence with the ID to ensure the stability of retrieval results; 2) The key information of the documents should be stored in the index as completely as possible to avoid the loss of information related to the query and thus affecting the retrieval results; 3) The model should be able to output documents in decreasing order of their relevance to the query. These three abilities are summarized as \textbf{exclusivity}, \textbf{completeness}, and \textbf{relevance ordering}. Our analytical experiments indicate that the currently available DSI models do not fully meet the requirements of a general end-to-end indexing-retrieval model, which limits their conditions of use and significantly reduces their effectiveness, especially when compared to the state-of-the-art dense retrieval models, which, in contrast, is shown to better meet those requirements. 


To this end, we investigate whether DSI models can be better trained to improve retrieval abilities while maintaining their simple structure and low storage cost. Specifically, we propose utilizing a dense retrieval approach to provide effective supervision signals for training DSI models. To enhance \textit{exclusivity} and \textit{completeness} of DSI, we propose to improve the document representation to capture information from different granularities and filter key information using the document representation encoded by the dense retrieval model. To improve the ability to discriminate the relevance degree of different documents of DSI, we propose a new distillation-based training approach. By explicitly modeling the connections between documents, the model is able to reduce the randomness of the output results and improve retrieval performance, especially on datasets with deep pool annotations. 


Major contributions of this paper are tri-fold. 1) An empirical analysis of basic IR abilities indicates the potential weaknesses of existing DSI approaches. 2) Based on the analysis above, we propose a multi-task distillation approach to improve the effectiveness of DSI by learning from dense retrieval while keeping its advantages. 3) Further evaluation shows that our approach substantially improves retrieval effectiveness of DSI-QG.


\section{Empirical Analysis} \label{mod_ana}
In this section, we conduct an empirical analysis to examine to what extent the DSI framework satisfies the basic requirements of a functioning IR model. 
Specifically, we summarize \textit{exclusivity}, \textit{completeness}, and \textit{relevance ordering}, as three essential abilities required for an IR framework, as defined below. 
\tcolor{While acknowledging that there are certain retrieval problems that do not require an ordered list, it is crucial to emphasize that our research specifically focuses on ranked retrieval, which inherently involves non-boolean ranking.}
The notions used throughout this paper are listed in Table~\ref{tab:notation} in Appendix~\ref{sec:notation}.

\begin{figure*}[tbp]
  \centering
  \setlength{\belowcaptionskip}{-0.4cm}
  \subfloat[
  Exclusivity\label{fig:exclusivity}] {\includegraphics[width=0.3\textwidth]{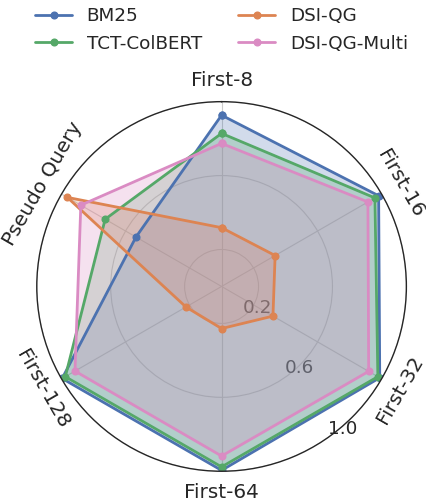}}
 \hfill 	
  \subfloat[
  Completeness\label{fig:completene}
  ]{\includegraphics[width=0.3\textwidth]{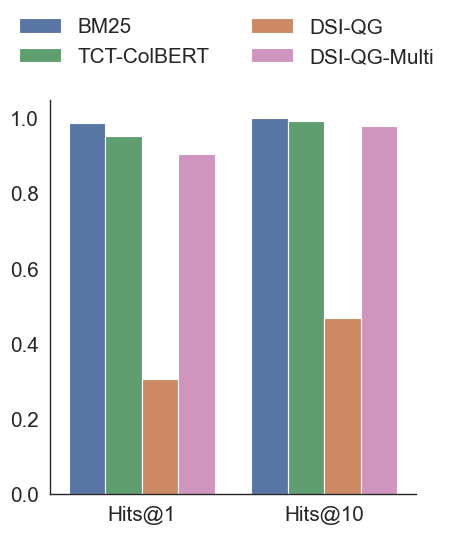}}
 \hfill	
  \subfloat[Relevance ordering\label{fig:relevance}]{\includegraphics[width=0.29\textwidth]{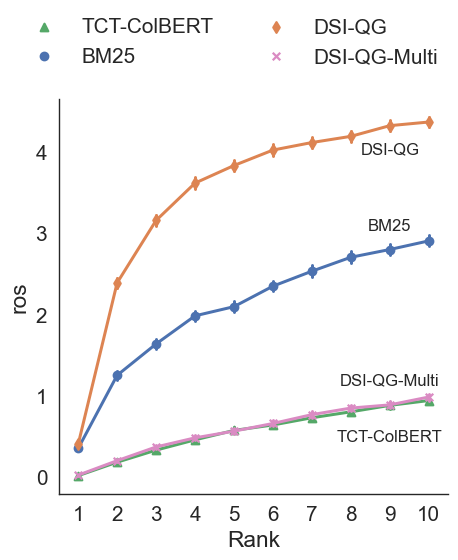}}
\\	
	\caption{Results of empirical analysis. (a) Hits@1 by querying different retrieval models with the first-\textit{k} tokens or a random pseudo query generated from the passage. \textbf{Exclusivity} shows the ability to retrieve a document by using its own content as a query. (b) Hits@1 and Hits@10 by querying different retrieval models with the key fragments within individual documents. \textbf{Completeness} shows the ability to retrieve a document by using its key content as a query. (c) The relevance ordering scores (y-axis) at different positions of the retrieval results (x-axis) for different retrieval methods. A higher \textbf{relevance ordering} score indicates a lower ability of the model to distinguish relevant documents from random ones. As DSI-QG-Multi (proposed in Section \ref{subsubsec:multi}) is trained using TCT-ColBERT's output as supervision signals, their curves (the bottom two) are almost indistinguishable.}
       \label{fig:analysis}
\end{figure*}

\subsection{Definitions}

\textbf{Exclusivity} refers to the uniqueness of documents in an IR system, i.e., the one-to-one correspondence between document content and its identifier, which determines the extent to which a retrieval framework can distinguish different documents in a collection. 
\tcolor{Although it is typical for document content and identifiers to have a one-to-one mapping, it is important to note that certain collections may contain duplicate documents, particularly in real-world scenarios. While exclusivity is primarily intended for stable experimental evaluations and reproducibility, it is not an absolute requirement for an index and is contingent upon the specific document collection. For the sake of simplicity and considering the majority of cases, we assume that the documents utilized in our experiments are not duplicated.}
In actual implementation, DSI is only trained for a one-way mapping (i.e., from document representation to document identifier)~\citep{DBLP:journals/corr/abs-2202-06991, DBLP:journals/corr/abs-2206-10128,DBLP:journals/corr/abs-2206-02743, DBLP:journals/corr/abs-2208-09257}. Therefore, in our analysis, we examine the injective relationship for various models by testing the case of a unidirectional mapping from a document to its identifier.

\textbf{Completeness} is the ability of an index to retrieve all eligible documents in a collection \tcolor{based on a specific query or search intent}. This means that if a document has relevant content and meets the search criteria, then that document should appear in the search results. To obtain more comprehensive search results, the index structure needs to store as complete document information as possible (e.g., sparse search indexes) or to maximize the identification and memory of more valuable key document information. In this context, the completeness of an index is narrowly defined as its ability to store essential information contained within a document \tcolor{that can be used to respond to a variety of queries}, taking into consideration the various differences in index structures and practical application requirements. 

\textbf{Relevance Ordering} is an essential ability of non-boolean IR models which outputs a sorted list of documents in decreasing order of relevance. 
For two documents $d_i$ and $d_j$ that both appear in the result list of a one-time retrieval operation $\operatorname{R}(q_t, \operatorname{M})$ on query $q_t$ based on a certain retrieval paradigm $\operatorname{M}$, relevance ordering can be described as:
\begin{align}
    d_j \leq_{q_t} d_i, \ & \text{if}\ \operatorname{s}(q_t, d_j) \leq \operatorname{s}(q_t, d_i), \notag \\ 
    & \text{for} \ \forall d_i, d_j \in R(q_t, \operatorname{M})\; 
\end{align}
\noindent where $d_j \leq_{q_t} d_i$ means a binary relation that $d_i$ is more relevant to $q_t$ than $d_j$, and $\operatorname{s}(q_t, d_*)$ is the predicted relevance score of $d_*$, which is in the top-\textit{k} recall list $(id^1, id^2, ..., id^k)$ given by $\operatorname{R}(q_t, \operatorname{M})$. 

For $d_i$ in the top-\textit{k} recall list of $\operatorname{R}(q_t, \operatorname{M})$ and $d_o$ in the index pool but not in the top-\textit{k} list, relevance ordering can be given as:
\begin{align}
    d_o \leq_{q_t} d_i,\ & \text{for} \ \forall d_i \in R(q_t, \operatorname{M})\notag \\& \text{and}\;\forall d_o \in \mathcal{D} \setminus R(q_t, \operatorname{M})
\end{align}
where $\mathcal{D} \setminus R(q_t, \operatorname{M})$ is the absolute complement of the set $R(q_t, \operatorname{M})$ in the indexing set $\mathcal{D}$.

\subsection{Analysis Methodology}
To assess the \textbf{exclusivity} of various models, we query different models with a cut-off of the first fragment from the original document, or pseudo-queries derived from the original documents, and assess their ability to retrieve the corresponding documents as the highest-ranked retrieval outcomes. Following the DSI-QG setup~\citep{DBLP:journals/corr/abs-2206-10128}, a series of experiments are conducted utilizing the training set of MS MARCO 100k~\citep{DBLP:conf/nips/NguyenRSGTMD16,DBLP:journals/corr/abs-2206-10128,DBLP:journals/corr/abs-2208-09257} to evaluate the exclusivity of various models. MS MARCO 100k is a subset of the MS MARCO passage dataset, which has been used in quite a few recent studies on DSI~\citep{DBLP:journals/corr/abs-2206-10128,DBLP:journals/corr/abs-2208-09257}.
The length of first-\textit{k} cut-off from the original document is varied within {16, 32, 64, 128}. The pseudo queries are generated by docT5query~\citep{nogueira2019doc2query},  based on the full text of the passage and followed the same procedures as the DSI-QG training data generation process~\citep{DBLP:journals/corr/abs-2206-10128}.
 
To measure the \textbf{completeness} of information stored in an index, we randomly select 10k documents from the training set of MS MARCO 100k~\citep{DBLP:conf/nips/NguyenRSGTMD16,DBLP:journals/corr/abs-2206-10128,DBLP:journals/corr/abs-2208-09257} and their corresponding queries. 
\tcolor{Assuming that the BERT reranker with cross-attention~\citep{DBLP:journals/corr/abs-1901-04085} behaves in a manner that closely approximates the ranking preferences of human annotators for documents, we employ a strategy to identify the most pertinent segments within each document.
To achieve this, we divide the documents into equally-sized chunks with overlaps.}
Each chunk is then scored by the BERT reranker for its relevance to the query.
\tcolor{It is important to acknowledge that it can be challenging to fully determine all of the key components of a document due to the lack of human annotations. Therefore, it is assumed that the best-scored chunks to the user queries constitute a subset of the document's essential content. This subset is then used as queries in various retrieval models, and their ability to accurately recall the corresponding document is evaluated as a measure of completeness.}

To assess the \textbf{relevance ordering} ability, in the MS MARCO passage dataset 
\citep{DBLP:conf/nips/NguyenRSGTMD16}, we employ the same BERT reranker as above. We randomly select 10k queries from the training set of MS MARCO 100k and use each retrieval model to predict the top 10 documents for each query. 
The binary group $(q,d_m)$ comprising of $q$ and the document $d_m$ in the result list is incorporated into a set referred to as $\mathcal{S}$. 
In the context of a retrieval model $\operatorname{M}$, the ranking of document $d_m$ in the result list returned for a query $q$ is represented by $r_{\operatorname{M}}(q, d_m)$, where $d_m\in \mathcal{S}$.
Additionally, a random selection of 10 documents from the dataset is made for $q$. For each randomly selected document $d_r$, the binary group $(q,d_r)$, is incorporated into a set referred to as $\mathcal{T}$. 
Subsequently, the BERT reranker is utilized to evaluate the relevance of the documents in the sets $\mathcal{S}$ and $\mathcal{T}$ to their corresponding queries, resulting in the ranking of document $d_i$ in the combined result list as $r_{BERT}(q, d_i)$. 

To evaluate the performance of a given model $\operatorname{M}$, 
we define a \emph{relevance ordering score} of $\operatorname{M}$ at the $p$th position in the result list returned for $q$ as: 
\begin{align}
    ros_q(\operatorname{M},p) = & ct(r_{BERT}(q, d_r)< r_{BERT}(q, d_m)),\notag \\
    & \text{for}\ d_r\in \mathcal{T}\ \text{and}\ r_{\operatorname{M}}(q, d_m)=p
\end{align}
where $ct(\cdot)$ is a counting function that determines the number of randomly selected documents that possess a higher relevance score given by BERT than $d_m$.
Upon conducting an average computation across all queries, the relevance ordering score for the retrieval method $\operatorname{M}$ at position $p$ is obtained:
\begin{equation}
    ros(\operatorname{M},p)=\mathop{Mean}\limits_{q\in Q}(ros_q(\operatorname{M},p))
\end{equation}
It is important to note that, as we assume that none of the random documents are relevant to the query if a significant proportion of these documents exhibit higher scores than those returned by the model, this serves as an indication that the model is returning less relevant documents. Models from different IR paradigms, including DSI-QG~\citep{DBLP:journals/corr/abs-2206-10128} which improves upon DSI, the BM25 sparse retrieval model~\citep{DBLP:journals/ftir/RobertsonZ09}, and the state-of-the-art dense retrieval model TCT-ColBERT (V2)~\citep{lin-etal-2021-batch}, are involved in our analysis.

\subsection{Analysis Results}

For \textbf{exclusivity}, Figure~\ref{fig:exclusivity} reports Hits@1 to evaluate the ability of different models to recall the document ID when using either a cut-off of the initial text from the document or pseudo-queries as input. The results indicate that both BM25 and TCT-ColBERT exhibit exceptional performance in accurately recalling the corresponding document ID when being queried with the original text cut-off, and the accuracy increases with the length of the cut-off. 
However, DSI-QG can recall the correct document ID for pseudo-queries, but struggles to do so when queried with text cut-off from the document itself.
This indicates that the specificity of DSI-QG is limited mostly to mapping from pseudo-queries to the associated document IDs, and lacks the ability to determine the document to which a segment of the primary text pertains, which could negatively impact its retrieval effectiveness.

For \textbf{completeness}, Figure\ref{fig:completene} shows the effectiveness of different retrieval methods in identifying important information in a document. Both BM25 and TCT-ColBERT have a high probability of accurately returning the correct document, with a probability above 95\% when considering the top result, and over 99\% when looking at the top 10 results. This suggests that both sparse and dense indexing methods effectively retain crucial content of the original document in relation to the user's query. In contrast, DSI-QG lacks proficiency in identifying the correct document, indicating that under the current training methodology, the differentiable search index may fail to capture important information, resulting in suboptimal performance.

Figure~\ref{fig:relevance} plots the \textbf{Relevance Ordering} score (y-axis) at each ranking position (x-axis). As the ranking progresses further down, the difference between the returned documents of and randomly selected documents becomes increasingly insignificant, which aligns with expectation. However, the second-ranked document of DSI-QG shows a significant decline in ros compared to TCT-ColBERT and BM25. As the ranking lowers, the differentiation between the returned documents of DSI-QG and randomly selected documents becomes increasingly indistinguishable. By the tenth document, approximately half of the random documents score higher, indicating a near-random distribution of the results of DSI-QG at that position.
Based on the observations above, we propose to improve DSI-QG by a multi-task distillation approach.

\section{Model Improvement}
\subsection{Method}
In this section, we propose to improve the DSI-QG framework in order to gain enhanced abilities in terms of exclusivity, completeness, and relevance ordering, with the help of supervision signals from dense retrieval models. 
Specifically, the dense models are utilized to search through text fragments comprising all indexed documents and pseudo-query texts generated from these fragments. 
The text fragments that effectively recall the original documents are then selected and added to the training data to provide a more unique and comprehensive collection of information. 
Furthermore, a new distillation-based model training task, utilizing document IDs recalled by the dense model as a supervision signal, is also proposed to address the semantic gap between short input text and single document ID label. 
Finally, the training tasks of the DSI model are reclassified in accordance with their respective task characteristics. By utilizing this classification, the model is trained with a multi-task setting, allowing for the improvement of various abilities through the aid of different tasks.

\subsubsection{Training Data Construction} \label{sec:data_select}

Current DSI methods mostly largely rely on supervised learning from training data to guarantee effectiveness. 
Due to the inevitable bottleneck of model memory against the sheer size of document corpus, DSI-QG, along with other DSI models like NCI~\citep{DBLP:journals/corr/abs-2206-02743} and Ultron~\citep{DBLP:journals/corr/abs-2208-09257}, resort to memorizing the much shorter pseudo queries, hence the gap between the query and the document representations. To this end, in order to improve the exclusivity of DSI-QG, we employ a two-phase procedure for constructing training data. The first step involves dividing each document $d_i$ in the set $\mathcal{D}$ of $n$ documents into equal-length segments with overlaps $\mathcal{O}_i=\{d_i^1,d_i^2,\cdots, d_i^m\}$. 
The resulting segments of all documents after this process form the set $\mathcal{O}=\{\mathcal{O}_1,\mathcal{O}_2,\cdots, \mathcal{O}_n\}$.
Subsequently, for each text segment $d_i^j$, in $\mathcal{O}$, we utilize docT5query~\citep{nogueira2019doc2query} to generate a pseudo query $pq_i^j$. 
Similar to the construction of $\mathcal{O}$, the pseudo query set $\mathcal{P}_i=\{pq_i^1,pq_i^2,\cdots, pq_i^m\}$ is obtained for each document, and $\mathcal{P}_i$ constitute the pseudo query set $\mathcal{P}$ of the entire dataset.
A combination of these two sets, $\mathcal{U}=\mathcal{O}\cup\mathcal{P}$, is expected to ensure a satisfactory level of information retention and effectively align the document text with the query.

Utilizing the set $\mathcal{U}$ for training may present challenges, such as difficulty in memorization due to a large number of text fragments and the inclusion of excessive irrelevant information. To mitigate these issues, filtering of $\mathcal{U}$ is implemented to optimize retention of crucial information.
As previous analysis in section~\ref{mod_ana} demonstrates that the dense retrieval model effectively preserves textual information in its representation, we employ the dense method $\operatorname{M}$ for further processing of  $\mathcal{U}$ by selectively filtering key content from it.
The process involves inputting individual text fragment $t$, originating from the document of $id_t$ in $\mathcal{U}$ as queries into the model, and obtaining a list, denoted as $\operatorname{R_k}(t, \operatorname{M})=(id^1, id^2,\cdots, id^k)$, of the top $k$ document IDs returned by the model. If $id_t$ of the original text fragment is present within  $\operatorname{R_k}(t, \operatorname{M})$, it is determined that the fragment $t$ possessed key information relevant to the original document, and is included in the corpus $\mathcal{T'}$ for the model training.

In our implementation, the value of $k$ is set to 1 for the original text segments and to 5 for the pseudo queries. 
$\mathcal{T'}$ was included in the initial training data for DSI-QG along with the original queries to ensure a valid comparison of results, resulting in the final training data $\mathcal{T}$.

\subsubsection{Explicit Modeling of Relevance}
Our previous analysis has shown that in order to improve the accuracy of retrieval results, the differentiable search index model should prioritize the enhancement of its relevance weighting capabilities in addition to its current unidirectional mapping to document IDs.
The current training approach for the DSI-QG model, which maps brief text to a single document identifier, is constrained by the limited amount of information present in the brief query text, which leads to the gap between the pseudo queries and document contents.
Moreover, the filtering of training data may cause the loss of information for certain documents, as the model would not be able to learn the relevant information of omitted documents if their text is not included.

In light of the aforementioned considerations, a distillation-based training protocol for DSI is proposed in order to explicitly model the relevance of various documents to a given query. 
For text fragment $t$ from the training set $\mathcal{T}$, the dense retrieval model $\operatorname{M}$ is utilized to query the corpus with $t$. The result list of the top $f$ document IDs returned by the model, denoted as  $\operatorname{R_f}(t, \operatorname{M})=(id^1, id^2,\cdots, id^f)$, in conjunction with the identifier $id_t$ of the document to which $t$ belongs, serves as the supervision signals for the entire distillation task. Formally, the supervision signal for text segment $t$ in the distillation task is defined as:
\begin{align}
    supsig^{dis}_t &= (id_t, \operatorname{R_f}(t, \operatorname{M})) \\ \notag
    &= (id_t, id^1, id^2,\cdots, id^f) 
\end{align}
During the implementation of training procedures, we set $f$ to 10 and utilize commas as a means of combining all identifiers into a cohesive string format, and the training objective of this task is:
\begin{equation}
    L_{dis}(\theta)=\sum\limits_{t\in \mathcal{T}}\text{log} \ p(supsig^{dis}_t|\operatorname{M}(t), \theta)
\end{equation}
\noindent where $p(supsig^{dis}_t|\operatorname{M}(t), \theta)$ denotes the probability of generating the supervised string given $t$ as the input of the model.

The distillation process is expected to enable the model to acquire knowledge from the precise ID correspondence with the text, and associate the (pseudo) query to the list of document identifiers that the dense model deems most pertinent. 
It is our contention that learning the relevance relationship between the input text and different documents could improve the model's ability of relevance ordering, resulting in more relevant documents appearing at the top of the list of results.
However, prior to this endeavor, the DSI approaches have been evaluated on datasets 
that come with only shallow relevance annotations, where a query usually has a single labeled relevant document. To validate the model in the context of deeply annotated data, we have constructed an MS MARCO 300k dataset, comprised of TREC DL19~\citep{DBLP:journals/corr/abs-2003-07820} and 20~\citep{DBLP:journals/corr/abs-2102-07662} data based on the MS MARCO passage set with 80 relevant documents per query on average. 

\subsubsection{End-to-end Multi-task Training}\label{subsubsec:multi}


Remind that in DSI~\citep{DBLP:journals/corr/abs-2202-06991}, the training task is divided into two sub-tasks, indexing and retrieval, depending on whether the input text is an original text fragment or a query. 
Our proposed model utilizes a multi-task setup, with however redefined tasks.
Empirically, both sub-tasks of DSI in fact lead the model to learn the features of a single document, thus we uniformly attribute them to the indexing task. Therefore, we formally define the indexing task as:
\begin{equation}
    t \rightarrow id_t, \ \text{for} \ t \in \mathcal{T}
\end{equation}
where $\rightarrow$ indicates an injective function that maps a text segment to its corresponding identifier. During training, the loss function of the indexing task is:
\begin{equation}
    L_{index}(\theta)=\sum\limits_{t\in \mathcal{T}}\text{log}\ p(id_t|\operatorname{M}(t), \theta)
\end{equation}
\noindent which maximizes the probability of generating $id_t$ with $t$ as the input of the model, and $t$ can be a natural text fragment in any form.

In contrast, our newly proposed training task utilizes a list of document IDs as a supervision signal, which enables the model to explicitly learn the relationship of relevance between documents and queries, as well as the relationship of relevance between documents themselves. Therefore, we define this task as the retrieval task:
\begin{equation}
    t \rightarrow (id^1, id^2,\cdots, id^l), \ \text{for} \ t \in \mathcal{T}
\end{equation}
The objective of this task is to assign a text fragment $t$ to a list of identifiers, which can be derived from any method. Here, we use $supsig^{dis}_t$ as the list, thus the loss function for the retrieval task is: 
\begin{equation}
    L_{retvl}(\theta)=L_{dis}(\theta)
\end{equation}
During training, the model is randomly presented with either a single ID or a list of IDs as the supervision signal with equal probability, and a special symbol is appended to the  beginning of the query to indicate the task type. The loss function of the multi-task training can be written as:
\begin{equation}
    L_{multi}(\theta)=L_{index}(\theta) + L_{retvl}(\theta)
\end{equation}

In the following, for the DSI-QG model that only uses the newly constructed training data, we denote it as \textbf{DSI-QG-Merge}.
The model that only uses the distillation task for training is labeled as \textbf{DSI-QG-Distill}, and the model that uses the newly constructed training data for multi-task training, first training the index task and then training the distillation task, is labeled as \textbf{DSI-QG-M+D}. The model trained using all of the above improvements is denoted as \textbf{DSI-QG-Multi}.
\subsection{Evaluation Setup}

\textbf{Datasets and Metrics.}
We experiment on the MS MARCO~\citep{DBLP:conf/nips/NguyenRSGTMD16} and Natural Question (NQ)~\citep{DBLP:journals/tacl/KwiatkowskiPRCP19} datasets. Akin to prior art~\citep{DBLP:journals/corr/abs-2202-06991, DBLP:journals/corr/abs-2206-10128}, we employ a 100k subset of MS MARCO and the Dev queries with shallow annotations. 
%
Our constructed MS MARCO 300k is also used, which includes queries and documents from both the Dev set and TREC DL19~\citep{DBLP:journals/corr/abs-2003-07820} and 20~\citep{DBLP:journals/corr/abs-2102-07662}. 
For the NQ dataset, following the DSI experimental procedures~\citep{DBLP:journals/corr/abs-2202-06991}, we construct the NQ 320k dataset.
Akin to~\citep{DBLP:journals/corr/abs-2202-06991},
we report Hits@1 and Hits@10 on Dev set of the data. To accurately assess the retrieval effectiveness on datasets with deeper annotations such as TREC DL 19 and 20, NDCG@10 and P@10 are also reported. Statistical
significance for paired two-tailed t-test is reported.

\textbf{Baselines.} We evaluate against the following baselines: the Pyserini implementation of \textbf{BM25}~\citep{Lin_etal_SIGIR2021_Pyserini}, and the original \textbf{DSI}~\citep{DBLP:journals/corr/abs-2202-06991}, \textbf{DSI-QG}~\citep{DBLP:journals/corr/abs-2206-10128}, and the recently proposed \textbf{NCI}~\citep{DBLP:journals/corr/abs-2206-02743}. The state-of-the-art dense retrieval model \textbf{TCT-ColBERT (V2)}~\citep{lin-etal-2021-batch} and \textbf{SEAL} based on generative retrieval~\citep{bevilacqua2022autoregressive} are also included. Further information about the data and baselines can be found in Appendix~\ref{sec:appendix}.

\textbf{Implementation Details.}
Following the DSI and DSI-QG settings~\citep{DBLP:journals/corr/abs-2206-10128}, the models are initialized using standard pre-trained T5 models~\citep{DBLP:journals/corr/abs-1910-10683}.
The T5-Base and T5-Large models are trained with a batch size of 128, the learning rate is  5e-4, and the maximum number of training steps is determined based on the scale of training data, with options among \{1M,2M,3M\}.
All pseudo-queries are generated by docT5query~\citep{nogueira2019doc2query} based on T5-Large. For our proposed method, we adhered to the DSI approach~\citep{DBLP:journals/corr/abs-2202-06991} to generate semantic IDs based on the dense representation of TCT-ColBERT. 
Following DSI-QG \citep{DBLP:journals/corr/abs-2206-10128}, all DSI models, with the exception of NCI, are trained using Naive String Docids if there are no extra specifications.
We plan to make our code and data available to public. 
\subsection{Evaluation Results}

\begin{table*}[tbh]
\centering
\setlength{\belowcaptionskip}{-0.3cm}
\resizebox{0.85\textwidth}{!}{
\begin{tabular}{l|l|llllll}
\toprule
\multirow{2}{*}{Methods} &
  \multirow{2}{*}{Model Size/ Task} &
  \multicolumn{2}{c}{MS Marco 100k Dev} &
  \multicolumn{2}{c}{MS Marco 300k Dev} &
  \multicolumn{2}{c}{NQ 320k} \\  
 &
   &
  Hits@1 &
  \multicolumn{1}{l}{Hits@10} &
  \multicolumn{1}{l}{Hits@1} &
  \multicolumn{1}{l}{Hits@10} &
  \multicolumn{1}{l}{Hits@1} &
  Hits@10 \\ \hline
BM25           &   -/-    & 0.5398  & 0.8295 & 0.4404  & 0.7417  &  0.0834  & 0.3336 \\
TCT-ColBERT &
  Base/ - &
  {\ul 0.7074} &
  {\ul 0.9506} &
  {\ul 0.6032} &
  {\ul 0.9107} &
  {\ul 0.2411} &
  {\ul 0.7197} \\ \hline
DSI            & Base/ Index  & 0.0292  & 0.0682 & 0.0218  & 0.0543 & 0.0008  & 0.0083 \\
DSI            & Large/ Index & 0.0874  & 0.1948 & 0.0214 &  0.0652   & 0.0057  & 0.0493 \\
SEAL            & Base/ -  & 0.2802  & 0.6219 & 0.2322  & 0.5818 & 0.1377  & 0.5679 \\
DSI-QG         & Base/ Index  & 0.6085  & 0.8026 & 0.5123  & 0.7703 & 0.2193  & 0.5180 \\
DSI-QG         & Large/ Index & 0.6182  & 0.8024 & 0.5188 & 0.7589  & 0.2223  & 0.5189  \\
NCI \emph{(sem)}            & Base/ Index  & 0.6133  & 0.8670 & 0.5289  & 0.8244  & 0.2120   & 0.6999 \\
\hline
DSI-QG-Multi   & Base/ Index  & 0.6711$^{\dagger}$  & \textbf{0.9208}$^{\dagger}$ & 0.5390  & 0.8726 $^{\dagger}$ & 0.2190 & 0.6138       \\
DSI-QG-Multi   & Base/ Retrv  & 0.6711$^{\dagger}$  & 0.9143$^{\dagger}$ & 0.5219  & 0.8480$^{\dagger}$ & 0.2162 & 0.7003       \\
DSI-QG-Multi\emph{(sem)}   & Base/ Index &
  0.6626$^{\dagger}$ &
  0.9115$^{\dagger}$ &
  0.5615$^{\dagger}$ &
  0.8801$^{\dagger}$ &
  0.2390$^{\dagger}$ & \textbf{0.7202}$^{\dagger}$  \\
DSI-QG-Multi\emph{(sem)}   & Base/ Retrv &
   \textbf{0.6739}$^{\dagger}$ &
   0.9192$^{\dagger}$ &
   0.5589$^{\dagger}$ &
   0.8659$^{\dagger}$ &
  \textbf{0.2392 }$^{\dagger}$ &  0.7135$^{\dagger}$  \\
DSI-QG-Multi   & Large/ Index & 0.6625$^{\dagger}$ &  0.9130$^{\dagger}$ & \textbf{0.5746}$^{\dagger}$ & \textbf{0.8868}$^{\dagger}$ & 0.2206$^{\dagger}$ & 0.6304   \\ 
DSI-QG-Multi   & Large/ Retrv & 0.6593$^{\dagger}$ & 0.9138$^{\dagger}$  & 0.5741$^{\dagger}$ & 0.8754$^{\dagger}$ & 0.2285$^{\dagger}$ &  0.7110$^{\dagger}$ \\\bottomrule
\end{tabular}}
\caption{Evaluation results with shallow annotations. Method names suffixed with \emph{(sem)} indicates that the semantic IDs are used in the training process. Statistical significance at 0.05 relative to NCI is marked by $\dagger$.}
\label{tab:main_tab}
\end{table*}

\begin{table}[tbh]
\setlength{\belowcaptionskip}{-0.4cm}
\resizebox{\linewidth}{!}{
\begin{tabular}{@{}l|llll@{}}
\toprule
\multirow{2}{*}{Methods} & \multicolumn{2}{c}{DL19}         & \multicolumn{2}{c}{DL20}          \\ 
                        & NDCG@10         & P@10            & NDCG@10         & P@10            \\ \hline
BM25                    & 0.6843          & 0.8837         &  0.6873         &  0.4852               \\
TCT-ColBERT            & {\ul 0.7977}    & {\ul 0.9279}    & {\ul 0.8012}    & {\ul 0.6315}    \\ \hline
DSI                     & 0.2156          & 0.2209          & 0.1907          & 0.1167          \\
DSI-QG                  & 0.6922          & 0.8256          & 0.7348          & 0.5630          \\ 
NCI                     & 0.6725          & 0.8419          & 0.7127          & 0.5407          \\\hline
DSI-QG-Merge            & 0.7215$^{\dagger}$          & 0.8767$^{\dagger\ddagger}$          & 0.7501$^{\dagger}$          & 0.5815$^{\dagger}$          \\
DSI-QG-Distill          & 0.7836$^{\dagger\ddagger}$          & 0.9140$^{\dagger\ddagger}$          & 0.7801$^{\dagger\ddagger}$          & 0.6056$^{\dagger\ddagger}$          \\
DSI-QG-D+M              & 0.7838$^{\dagger\ddagger}$          & 0.9209$^{\dagger\ddagger}$          & 0.7851$^{\dagger\ddagger}$          & 0.6074$^{\dagger\ddagger}$          \\
DSI-QG-Multi            & \textbf{0.7920}$^{\dagger\ddagger}$ & \textbf{0.9279}$^{\dagger\ddagger}$ & \textbf{0.7983}$^{\dagger\ddagger}$ & \textbf{0.6278}$^{\dagger\ddagger}$ \\ \bottomrule
\end{tabular}}
\caption{Results on TREC DL datasets with deeper annotations. DSI-QG-D+M and DSI-QG-Multi are evaluated on the retrieval task while others are evaluated on the indexing task. Statistical significance at 0.05 relative to NCI or DSI-QG is marked by $\dagger$ or $\ddagger$.}
\label{tab:dl_tab}
\end{table}

\textbf{The enhanced DSI-QG variants outperform the DSI baselines.}
Results on datasets with shallow annotations are presented in Table\ref{tab:main_tab}. 
The proposed method of multi-task training, DSI-QG-Multi, demonstrates a notable enhancement in comparison to existing DSI models, with statiscally significant improvement reported on three datasets.
Additionally, our experiments demonstrate that scaling up the model size has limited impact on the original DSI-QG, whereas the our adjustment to the training data and training tasks allow model to obtain better results on larger models as the data size increases, as is evident on the MS MARCO 300k data. Furthermore, our utilization of multi-task training enables the selection of different subtask settings for prediction, with the indexing task yielding the best results for the MS MARCO dataset, and the retrieval setting producing better results for the NQ dataset, possibly due to the need to rely on different types of information for relevant documents for different datasets. 

\textbf{The proposed training approach facilitates the three aforementioned abilities of DSI-QG.}
The effectiveness of the proposed method in enhancing the basic abilities is evaluated through the DSI-QG-Multi model, for the three experiments described in Section~\ref{mod_ana}. The results of these experiments are included in Figure~\ref{fig:analysis}. The present study demonstrates that our improvement of DSI-QG results in significant enhancement in all three model abilities. 
Figure~\ref{fig:exclusivity} illustrates that the model is able to accurately return the corresponding IDs of the text of different lengths when the initial parts of documents are input, and its ability to identify documents corresponding to pseudo-queries is still maintained. 
Figure~\ref{fig:completene} further supports the validity of the model's stored information following optimized training, as it is able to accurately locate documents containing key content related to a given query. 
This demonstrates that our approach effectively picks up the unique and key information of the documents, and our model effectively encodes these contents in training, thus makes improvements on  exclusivity and completeness.

\textbf{Our proposed improvements has led to a significant reduction in the performance gap to TCT-ColBERT.} This is particularly encouraging as it suggests that our proposed DSI improvements are able to achieve comparable performance to the dense model on datasets with more complete annotations, as seen in Table \ref{tab:dl_tab}. Importantly, this is achieved while still maintaining the advantages of DSI, such as minimal storage cost and end-to-end retrievability. As depicted in Figure~\ref{fig:relevance}, the distribution of the top 10 documents returned by DSI-QG following optimized training exhibits a high degree of similarity to that of TCT-ColBERT. This demonstrates that the model is capable of effectively modeling the correlation order among documents through distillation-based training.  To further validate this conclusion, Table~\ref{tab:dl_tab} presents the effects of various models that were trained on MS MARCO 300k data on the TREC DL19 and DL20 query sets. The data demonstrate that our method consistently outperforms the original training method by at least 8.6\% on deeply annotated data, thus achieving a level of performance that is comparable to that of dense retrieval models.
\tcolor{While our DSI-QG-Multi model still slightly underperforms TCT-ColBERT on datasets with shallow annotations, it is important to note that our objective is to enhance the retrieval effectiveness of DSI models. Previous studies, such as NCI~\citep{DBLP:journals/corr/abs-2206-02743} and Ultron~\citep{DBLP:journals/corr/abs-2208-09257}, have indicated that DSI models can outperform dense retrieval models when trained on similar smaller datasets. In our experiments, TCT-ColBERT was trained on the entire MSMARCO dataset, which likely contributes to its superior performance. By distilling the ranking capabilities of TCT-ColBERT, we have achieved significant improvements.}



\textbf{Effects of Document Identifiers.}
To investigate the impact of various identifiers, we assign semantic IDs to documents utilizing the hierarchical clustering process as in DSI~\citep{DBLP:journals/corr/abs-2202-06991} in our training process. 
The results presented in table~\ref{tab:main_tab} indicate that the semantic IDs can further improve the retrieval on larger datasets (i.e., MS MARCO 300k and NQ).
For comparison of our model with other models in different tasks under semantic IDs, as well as the ablation analysis, please refer to Appendix~\ref{sec:identifier} \& \ref{sec:ablation}.

\section{Related Works}
The field of information retrieval has recently seen a surge in interest in generative retrieval models.
Examples of this approach include the docT5query~\citep{nogueira2019doc2query}, which trains T5 models to generate document-related queries and adds them to the original documents for index construction, and SEAL~\citep{DBLP:journals/corr/abs-2204-10628}, which generates text fragments that the target document may contain and uses FM-Index~\citep{DBLP:conf/focs/FerraginaM00} for retrieval.
GENRE~\citep{DBLP:conf/iclr/CaoI0P21}  utilizes BART~\citep{DBLP:conf/acl/LewisLGGMLSZ20} for entity name generation in entity retrieval tasks.

Furthermore,~\citet{DBLP:journals/corr/abs-2202-06991} proposed the Differentiable Search Index (DSI), which explores various strategies for mapping string queries to document IDs based on T5~\citep{DBLP:journals/corr/abs-1910-10683}. This idea has been further developed in models such as DSI-QG~\citep{DBLP:journals/corr/abs-2206-10128}, NCI~\citep{DBLP:journals/corr/abs-2206-02743}, and Ultron~\citep{DBLP:journals/corr/abs-2208-09257}, which have all effectively improved the retrieval performance of generative retrieval models through the use of pseudo queries.
More specifically, DSI-QG~\citep{DBLP:journals/corr/abs-2206-10128} adapted the model for multilingual retrieval tasks, Ultron~\citep{DBLP:journals/corr/abs-2208-09257} attempted to incorporate hyperlink information into document IDs, and NCI~\citep{DBLP:journals/corr/abs-2206-02743} employed richer data, more fine-grained semantic ID mapping, and a novel decoder structure to obtain the best retrieval results.
Differentiable indexing models have also been applied to a wide range of tasks such as knowledge-intensive language tasks~\citep{DBLP:conf/cikm/ChenZG0FC22}, long sequence retireval~\citep{DBLP:journals/corr/abs-2204-13596} and QA tasks~\citep{DBLP:journals/corr/abs-2209-10063}. 
Additionally, efforts have been made to improve the memory capability~\citep{DBLP:journals/corr/abs-2212-09744} and decoding capability~\citep{DBLP:journals/corr/abs-2210-02068} of these models. This paper aims to gain a deeper understanding of differentiable indexing models as a retrieval method and proposes a multi-task distillation approach to improve their performance.



\section{Conclusions}
We summarized three essential abilities of a functional non-Boolean IR framework. Through an empirical analysis of these abilities, we identified potential weaknesses in existing DSI approaches. To address these weaknesses, we propose a multi-task distillation approach to enhance the effectiveness of DSI by learning from dense retrieval while preserving the advantages of DSI, such as minimal storage cost and end-to-end retrievability. Our evaluation results indicate that our proposed approach improves the three abilities of DSI and, as a result, its retrieval effectiveness, particularly on data with more comprehensive human annotations.
\tcolor{There are several directions to explore. Firstly, the three capabilities identified in our empirical analysis, although currently applied to DSI, can be further extended to analyze statistical retrieval methods as well. Additionally, investigating the trade-off between these three capabilities would be an interesting avenue for future research. Performing more comprehensive experiments to ascertain the effectiveness of retrieval models that rely on expanded queries for downstream tasks would yield valuable insights.}

\clearpage
\section*{Limitations}
In this study, we primarily focus on the examination and experimentation of the DSI-QG model, with plans to expand our research to include more recent models that utilize differentiable search indexing, such as the NCI model. 
\tcolor{While our approach has demonstrated effective improvements in DSI retrieval outcomes, and both TCT-ColBERT and our proposed DSI-QG-Multi performed well in our empirical analysis concerning relevance ordering, we cannot dismiss the possibility that these favorable results may be attributed to the extraction of insights from a specific BERT reranker model that shares similarities or correlations with the one used to define the desired ranking.}

Despite showing improvement over DSI-QG, our model remains slightly less effective than state-of-the-art dense retrieval methods such as TCT-ColBERT V2. 
\tcolor{Our approach offers advantages over dense retrieval models such as reduced storage and maintenance overhead, as DSI models do not require additional index structures for online use. Though index structures are utilized during the training phase of DSI-QG-Multi, the generated index structures are temporary in nature. }

Furthermore, due to the limitation of model memory, current research on DSI only experiments on a subset of the entire MS MARCO dataset or small dataset such as NQ. Therefore, an important future direction is to develop more efficient architectures to deal with the issue of memory bottleneck, for example, by using the current popular Large Language Models (LLM) or constructing aggregation structures for storing all information in hierarchical pieces.

\noindent\textbf{Acknowledgements.} This work is supported in part by the National Key Research and Development Program of China (No. 2020AAA0106400) and the National Natural Science Foundation of China (No. 62272439).


\bibliography{anthology,custom}
\bibliographystyle{acl_natbib}
\clearpage
\appendix

\section{Appendix}
\label{sec:appendix}

\subsection{Notation Table} \label{sec:notation}
The notations are listed in table~\ref{tab:notation}
.\begin{table}
  \centering
  \caption{Summary of notation.}
  \label{tab:notation}
  \resizebox{\linewidth}{!}{
  \begin{tabular}{@{}l|l@{}}
    \toprule
    \textbf{Symbol} & \textbf{Definition} \\ \midrule
    $\mathcal{D}=\{d_i\}$ & Set of documents stored in the index. \\ [0.8ex]
    $\mathcal{I}=\{id_i\}$ & Set of docids of documents in $\mathcal{D}$. \\ [0.8ex]
    $\mathcal{Q}=\{q_i\}$ & Set of user queries. \\ [0.8ex]
    $\operatorname{M}(d_i)$ & An retrieval function take  $d_i$ as input. \\ [0.8ex]
    $\operatorname{R}(q_i, \operatorname{M})$  & A retrieval operation for $q_i$ based on $\operatorname{M}$. \\ [0.8ex]
    $\operatorname{s}(q_i, d_i) $ & Relevance score of $q_i$ and $d_i$. \\ [0.8ex]
    $d_j \leq_{q} d_i$ & \begin{tabular}[c]{@{}l@{}} A binary relation that $d_i$ is more\\ relevant to $q$ than $d_j$ \end{tabular}\\ \bottomrule
  \end{tabular}}
\end{table}

\subsection{Datasets Details}
For MS MARCO 100k dataset, We adhere to the experimental design outlined in~\citep{DBLP:journals/corr/abs-2206-10128} by randomly selecting 93k passages from the dataset and incorporating all text from the validation set. 
For MS MARCO 300k dataset, we randomly select 293k passages from the dataset and incorporate all text from the validation set, as well as from TREC DL19 and 20. 
For NQ 320k dataset, we follow the DSI experimental procedures~\citep{DBLP:journals/corr/abs-2202-06991}, constructing the data consisting of approximately 200k passages and 8k validation set queries after pre-processing. 
\subsection{Baseline Details} \label{sec:bsln_detail}
The details of our baselines are as follows:
\begin{itemize}
    \item \textbf{BM25}~\citep{DBLP:journals/ftir/RobertsonZ09} is a classical sparse retrieval model that utilizes lexical weights. In this study, we employ a pyserini-based implementation~\citep{DBLP:conf/sigir/LinMLYPN21}, which leverages the Lucene~\citep{DBLP:conf/sigir/BialeckiMI12} as the underlying infrastructure. 
    \item \textbf{TCT-ColBERT (V2)}~\citep{lin-etal-2021-batch} is a state-of-the-art single-vector dense retrieval model that utilizes knowledge distillation and hard negative example sampling techniques. The model combines the performance of ColBERT~\citep{DBLP:journals/corr/abs-2004-12832} with the computational efficiency of a bi-encoder. The implementation of TCT-ColBERT is based on the Faiss vector index and is implemented through the pyserini library. This model is utilized to demonstrate the effectiveness of the current state-of-the-art model for retrieval and serves as a guide for training a differentiable search index. Note that the TCT-ColBERT model is exclusively trained utilizing the MS MARCO Passage dataset. Our analysis of the retrieval performance of various dense retrieval models on NQ data revealed that TCT-ColBERT displayed remarkable results despite not having been specifically trained on that dataset.
    \item \textbf{DSI}~\citep{DBLP:journals/corr/abs-2202-06991} is a T5-based approach for learning text-to-identifier mappings. Specifically, DSI defines the process of mapping original text to identifiers as an indexing task and the mapping of query text to identifiers as a retrieval task. This study reproduces the DSI model using the Naive String Docid and Semantic String Docid techniques, based on T5-base and T5-large architectures, utilizing open-source implementations\footnote{\url{https://github.com/ArvinZhuang/DSI-transformers}}. As access to the original training data of DSI is not available, we followed the settings of the open-source implementation to set the indexing and retrieval ratio to 1:1 for MS MARCO and 3:1 for NQ. 
    \item \textbf{DSI-QG}~\citep{DBLP:journals/corr/abs-2206-10128} improves upon DSI by incorporating pseudo-queries that are generated utilizing DocT5Query during the training process. Our research has involved reproducing the model on the MS MARCO 100k dataset, utilizing the available open-source code, and subsequently applying it to the MS MARCO 300k and NQ 320k datasets. This model serves as the baseline for our work but also serves as the primary focus for our investigations into potential improvements.
    \item \textbf{NCI}~\citep{DBLP:journals/corr/abs-2206-02743}, is a recently proposed state-of-the-art differentiable search indexing model that utilizes a variety of techniques to enhance its performance. These include the generation of semantic identifiers, the implementation of query generation strategies, and the utilization of a prefix-aware weight-adaptive decoder. Through the use of open-source implementation, the effectiveness of the model is validated using the three distinct datasets.
    \item \textbf{SEAL}~\citep{bevilacqua2022autoregressive} is a novel methodology that incorporates an autoregressive language model with a compressed full-text substring index. The implementation of this model utilizes BART and an external index known as the FM-index. We evaluate it on three datasets based on its official implementation.
\end{itemize}

\subsection{Impact of Document Identifier} \label{sec:identifier}
\begin{table*}[tbh]
\resizebox{\textwidth}{!}{
\begin{tabular}{@{}l|l|llllllllll@{}}
\hline
\multirow{3}{*}{Models} &
  \multirow{3}{*}{ID Type} &
  \multicolumn{2}{c|}{MSMarco 100k} &
  \multicolumn{6}{c|}{MSMarco 300k} &
  \multicolumn{2}{c}{NQ 320k} \\ \cline{3-12} 
 &
   &
  \multicolumn{2}{c|}{Dev} &
  \multicolumn{2}{c|}{Dev} &
  \multicolumn{2}{c|}{DL19} &
  \multicolumn{2}{c|}{DL20} &
  \multicolumn{2}{c}{Dev} \\ \cline{3-12} 
 &
   &
  Hits@1 &
  \multicolumn{1}{l|}{Hits@10} &
  \multicolumn{1}{l|}{Hits@1} &
  \multicolumn{1}{l|}{Hits@10} &
  \multicolumn{1}{l|}{Ndcg@10} &
  \multicolumn{1}{l|}{P@10} &
  \multicolumn{1}{l|}{Ndcg@10} &
  \multicolumn{1}{l|}{P@10} &
  \multicolumn{1}{l|}{Hits@1} &
  Hits@10 \\ \hline
  NCI      & Semantic      & 0.6133  & 0.8670 & 0.5289  & 0.8244 & 0.6725          & 0.8419          & 0.7127          & 0.5407  & 0.2120   & 0.6999 \\ \hline
\multirow{2}{*}{DSI} &
  Naive &
  0.0292 &
  0.0682 &
  0.0218 &
  0.0543 &
  0.2156 &
  0.2209 &
  0.1907 &
  0.1167 &
  0.0008 &
  0.0083 \\
 &
  Semantic &
  0.0957 &
  0.2597 &
  0.0602 &
  0.2054 &
  0.3607 &
  0.4279 &
  0.2749 &
  0.2000 &
  0.0212 &
  0.2370 \\
\multirow{2}{*}{DSI-QG} &
  Naive &
  0.6085 &
  0.8026 &
  0.5123 &
  0.7703 &
  0.6922 &
  0.8256 &
  0.7348 &
  0.5630 &
  0.2193 &
  0.5180 \\
 &
  Semantic &
  0.6103 &
  0.8109 &
  0.5252 &
  0.7971 &
  0.7022 &
  0.8512 &
  0.7317 &
  0.5759 &
  0.2287 &
  0.6252
   \\ \hline
\multirow{2}{*}{DSI-QG-Multi} &
  Naive &
  \textbf{0.6711}$^{\dagger}$ &
  \textbf{0.9208}$^{\dagger}$ &
  0.5390 &
  0.8726$^{\dagger}$ &
  0.7920$^{\dagger}$ &
  0.9279$^{\dagger}$ &
  \textbf{0.7983}$^{\dagger}$ &
  \textbf{0.6278}$^{\dagger}$ &
  0.2190 & 0.6138 
   \\
 &
  Semantic &
  0.6626$^{\dagger}$ &
  0.9115$^{\dagger}$ &
  \textbf{0.5615}$^{\dagger}$ &
  \textbf{0.8801}$^{\dagger}$ &
  \textbf{0.7961}$^{\dagger}$ &
  \textbf{0.9372}$^{\dagger}$ &
  0.7844$^{\dagger}$ &
  0.6019$^{\dagger}$ &
  \textbf{0.2390}$^{\dagger}$ & \textbf{0.7202}$^{\dagger}$ 
   \\ \hline
\end{tabular}}
\caption{Semantic ID Experiment Results. Results on the dev set of the three datasets are based on index task, and results on TREC DL are based on retrieval task. Statistical significance at 0.05 relative to NCI is marked by $\dagger$.}
\label{tab:semantic_tab}
\end{table*}

To investigate the impact of various identifiers, we assign semantic IDs to documents utilizing the hierarchical clustering process as in DSI~\citep{DBLP:journals/corr/abs-2202-06991} and have subsequently trained various models based on these new IDs. 
From Table~\ref{tab:semantic_tab}, it is observed that for DSI and DSI-QG, the utilization of semantic clustering-based document IDs results in improved effectiveness when compared to the use of Naive String IDs. For our proposed DSI-QG-Multi, the use of semantic IDs is effective on larger datasets, such as MS MARCO 300k and NQ320k. Across all three models, it is evident that semantic IDs are highly beneficial in enhancing Hits@10 on NQ.


\subsection{Impact of Different Factors} \label{sec:ablation}

\begin{table*}[tbh]
\centering
\resizebox{0.85\textwidth}{!}{
\begin{tabular}{l|l|llllll}
\toprule
\multirow{2}{*}{Methods} &
  \multirow{2}{*}{Model Size/ Task} &
  \multicolumn{2}{c}{MS Marco 100k Dev} &
  \multicolumn{2}{c}{MS Marco 300k Dev} &
  \multicolumn{2}{c}{NQ 320k} \\  
 &
   &
  Hits@1 &
  \multicolumn{1}{l}{Hits@10} &
  \multicolumn{1}{l}{Hits@1} &
  \multicolumn{1}{l}{Hits@10} &
  \multicolumn{1}{l}{Hits@1} &
  Hits@10 \\ \hline
DSI-QG         & Base/ Index  & 0.6085  & 0.8026 & 0.5123  & 0.7703 & 0.2193  & 0.5180 \\
NCI            & Base/ Index  & 0.6133  & 0.8670 & 0.5289  & 0.8244  & 0.2120   & 0.6999 \\
\hline
DSI-QG-Distill & Base/ Retrv  & 0.6095  & 0.8851$^{\dagger}$ & 0.4755  & 0.8143 & 0.2188 & 0.6535  \\
DSI-QG-Merge   & Base/ Index  & 0.6457$^{\dagger}$  & 0.8692 & 0.5493$^{\dagger}$  & 0.8302 & \textbf{0.2199}$^{\dagger}$      & 0.5957   \\
DSI-QG-D+M   & Base/ Retrv  & 0.6460$^{\dagger}$  & 0.9097$^{\dagger}$ & 0.5007  & 0.8471$^{\dagger}$ & 0.2163 &  0.6992 \\
DSI-QG-Multi   & Base/ Index  & \textbf{0.6711}$^{\dagger}$  & \textbf{0.9208}$^{\dagger}$ & \textbf{0.5390}  & \textbf{0.8726} $^{\dagger}$ & 0.2190 & 0.6138       \\
DSI-QG-Multi   & Base/ Retrv  & \textbf{0.6711}$^{\dagger}$  & 0.9143$^{\dagger}$ & 0.5219  & 0.8480$^{\dagger}$ & 0.2162 & \textbf{0.7003}       \\
\bottomrule
\end{tabular}}
\caption{Ablation study results with shallow annotations. Statistical significance at 0.05 relative to NCI is marked by $\dagger$.}
\label{tab:ab_tab}
\end{table*}

As shown in table~\ref{tab:ab_tab},the results of \emph{DSI-QG-Distill} reveal that the distillation task effectively improves the performance of the DSI-QG model on Hits@10, while having comparable performance on Hits@1, except for a drop on the MS MARCO 300k data.
This suggests that the model is able to recall more relevant documents for a given query in the 2nd to 10th positions of the returned list, which is in line with our objective of enhancing the relevance ordering capabilities of the model through distillation.
In comparison to the original DSI-QG, \emph{DSI-QG-Merge} exhibits a substantial enhancement in both Hits@1 and Hits@10. This illustrates the non-negligible impact that training data has on the overall performance of the model, thereby highlighting the importance of utilizing a more judicious and directed approach in the selection of training data.
When the tasks are trained independently, the result of \emph{DSI-QG-M+D} demonstrates an improvement in Hits@10 across all three datasets in comparison to prior single-task training. However, this improvement is less substantial than the gain achieved by \emph{DSI-QG-Multi}.

\end{document}